\documentclass[conference]{IEEEtran}

\IEEEoverridecommandlockouts
\usepackage{todonotes}
\usepackage{threeparttable}
\usepackage{cite}
\usepackage{amsmath,amssymb,amsfonts}
\usepackage{acronym}
\usepackage{algorithmic}
\usepackage{enumitem}
\usepackage[hidelinks]{hyperref}
\usepackage{tabularray}
\usepackage{graphicx}
\usepackage{textcomp}

\usepackage{tabularx}
\usepackage{multirow}
\usepackage{balance}
\usepackage{fancyhdr}
\renewcommand{\footrulewidth}{0.4pt}
\usepackage{siunitx}  
\usepackage{bbm}

\usepackage{hhline}
\usepackage{makecell} 
\usepackage[a4paper, total={184mm,239mm}]{geometry}
\def\BibTeX{{\rm B\kern-.05em{\sc i\kern-.025em b}\kern-.08em
    T\kern-.1667em\lower.7ex\hbox{E}\kern-.125emX}}

\usepackage[shortcuts,acronym]{glossaries}
\usepackage{fancyhdr}

\begin{document}

\makeatletter
\def\footnoterule{\kern-3\p@
  \hrule \@width 0.75in \kern 2.6\p@} 
\makeatother

\makeatletter
\newcommand{\linebreakand}{%
  \end{@IEEEauthorhalign}
  \hfill\mbox{}\par
  \mbox{}\hfill\begin{@IEEEauthorhalign}
}
\makeatother
\bstctlcite{IEEEexample:BSTcontrol}

\newcommand{\red}[1]{{\color{black}#1}}
\newcommand{\blue}[1]{{\color{black}#1}}
\newcommand{\orange}[1]{{\color{black}#1}}
\newcommand{\purple}[1]{{\color{black}#1}}

\title{Function Approximation Using Analog Building Blocks in Flexible Electronics}

\fancypagestyle{firstpage}{%
    \fancyhf{}  
    \renewcommand{\headrulewidth}{0pt}  
    \renewcommand{\footrulewidth}{0pt}
    \fancyhead[C]{Accepted for publication at the 26th International Symposium on Quality Electronic Design (ISQED'25), April 23-25, 2025.}
}

\author{
    \IEEEauthorblockN{
        Paula Carolina Lozano Duarte\IEEEauthorrefmark{1},
        Aradhana Dube\IEEEauthorrefmark{1},
        Georgios Zervakis\IEEEauthorrefmark{2},
        Mehdi Tahoori\IEEEauthorrefmark{1},
        Sani Nassif\IEEEauthorrefmark{3}
    }
    \IEEEauthorblockA{
        \IEEEauthorrefmark{1}Karlsruhe Institute of Technology,
        \IEEEauthorrefmark{2}University of Patras,
        \IEEEauthorrefmark{3}Radyalis LLC
    }
    \IEEEauthorblockA{
        \IEEEauthorrefmark{1}\{paula.duarte, aradhana.dube, mehdi.tahoori\}@kit.edu,\\
        \IEEEauthorrefmark{2}zervakis@ceid.upatras.gr,
        \IEEEauthorrefmark{3}srn@radyalis.com,
    }
}

\maketitle
\thispagestyle{firstpage} 

\renewcommand{\figureautorefname}{Fig.}
\renewcommand{\equationautorefname}{Eq.}
\renewcommand{\sectionautorefname}{Sec.}
\renewcommand{\subsectionautorefname}{Sec.}
\renewcommand{\subsubsectionautorefname}{Sec.}
\renewcommand{\tableautorefname}{Tab.}

\newcommand{\vx}{\ensuremath{\mathrm{\boldsymbol{x}}}}
\newcommand{\vy}{\ensuremath{\mathrm{\boldsymbol{y}}}}
\newcommand{\vw}{\ensuremath{\mathrm{\boldsymbol{w}}}}
\newcommand{\vg}{\ensuremath{\mathrm{\boldsymbol{g}}}}
\newcommand{\vW}{\ensuremath{\mathrm{\boldsymbol{W}}}}
\newcommand{\vV}{\ensuremath{\mathrm{\boldsymbol{V}}}}
\newcommand{\vP}{\ensuremath{\mathrm{\boldsymbol{P}}}}
\newcommand{\gmax}{\ensuremath{\mathrm{G_{max}}}}
\newcommand{\gmin}{\ensuremath{\mathrm{G_{min}}}}
\newcommand{\veta}{\ensuremath{\mathrm{\boldsymbol{\eta}}}}
\newcommand{\vq}{\ensuremath{\mathrm{\boldsymbol{q}}}}
\newcommand{\vTheta}{\ensuremath{\mathrm{\boldsymbol{\Theta}}}}

\begin{abstract}

\blue{Function approximation \purple{is crucial} in Flexible Electronics (FE), \purple{where applications demand efficient computational techniques within strict constraints on size, power, and performance.
Devices like wearables and compact sensors are constrained by their limited physical dimensions and energy capacity, making traditional digital function approximation challenging and hardware-demanding.}
This paper addresses function approximation in FE by proposing a \purple{systematic and generic} approach using a combination of Analog Building Blocks (ABBs) that perform basic mathematical operations such as addition, multiplication, and squaring.
These ABBs serve as the foundation for constructing splines, which are then employed in the creation of Kolmogorov-Arnold Networks (KANs), improving the approximation.  
The analog realization of KAN offers a promising alternative to digital solutions, providing significant hardware benefits, particularly in terms of area and power consumption. 
Our design achieves a 125$\times$ reduction in area and a 10.59\% power saving compared to a digital spline with 8-bit precision.
Results \purple{also} show that the analog design introduces an approximation error of up to 7.58\% due to both the design and parasitic elements.
Nevertheless, KANs are shown to be a viable candidate for function approximation in FE, with potential for further optimization to address the challenges of error reduction and \purple{hardware cost.}}
\end{abstract}

\vspace{-0.5ex}

\section{Introduction}\label{sec:intro}

\blue{
Flexible Electronics (FE) have emerged as a promising technology with the potential to revolutionize a wide range of applications, including wearable devices, sensors, and next-generation medical technologies~\cite{Gao:FlexibleWearableSensing, Heng:AM2022:FlexHumanMachInterfaces,gao:2016:flexsensor}. 
FE, based on materials like Indium Gallium Zinc Oxide (IGZO), offers advantages such as lightweight, low cost \purple{manufacturing}, and adaptable form factors, making it suitable for applications where traditional silicon-based electronics may fall short~\cite{Baruah:FabricationFE2023,Jeong:igzoperformance}. 
However, the integration of FE into practical, high-performance systems requires overcoming several inherent challenges, primarily the need for low power, small footprint, and efficient near-sensor or on-sensor processing~\cite{Arokia:feAdvantages2012,Lozano:aspdac25:BinCoDesign}.

One of the key limitations of FE is the larger area required for complex circuits compared to traditional silicon-based systems \purple{due to larger feature sizes}, which makes adopting digital processing techniques bulky and costly for certain applications. 
\purple{As FE applications demand greater functionality, digital hardware becomes increasingly challenging in terms of both cost and performance}~\cite{Baruah:FabricationFE2023,Arokia:feAdvantages2012,Ozer:Nature:2020,Sarpeshkar:AnalogvsDigital1998}.}

The ability to approximate functions efficiently is crucial for many FE applications, \purple{such as monitoring biological signals in wearables (e.g., calculating average heart rate or temperature) or detecting environmental parameters in compact IoT sensors (e.g., approximating pollution levels or humidity)~\cite{Gao:FlexibleWearableSensing,Arokia:feAdvantages2012,Haghi:wearableinhealthmonitoring2021}.
Function approximation enables systems to process signals in real-time while using minimal power and hardware. 
However, traditional digital function approximation methods can be too power-hungry and hardware-demanding for small-scale, low-cost FE applications.}
Kolmogorov-Arnold Networks (KANs) offer a powerful framework for function approximation, providing an efficient and systematic way to model complex relationships in data through the use of spline functions~\cite{Liu:KAN2024}. 
\purple{KANs offer the flexibility to approximate any given function, making them ideal for use in a variety of applications~\cite{Sidhart:fuctionapproxKAN2024}.
}

The key challenge, however, lies in efficiently realizing these approximations in hardware, \blue{particularly given the constraints of FE.} While digital implementations of KANs are straightforward, their associated power consumption and hardware costs often present significant barriers, and this has been studied as a co-optimization problem~\cite{VanDuy:KANCodesign:2024,Huang:KANAcceletaror:2024}.
\blue{Analog realizations of KANs must address issues such as maintaining precision, minimizing power consumption, and dealing with parasitic elements inherent to circuit designs.}

This work proposes a novel solution by using analog circuits to implement KANs for function approximation in FE.
\purple{The key feature of our approach is that it is systematic and generic, enabling analog approximations of any function through a standardized design process that brings structure and precision to analog function approximation.
As a result, our approach overcomes the limitations of traditional methods and provides a reliable way to implement analog function approximation in different applications.}
We present the development of Analog Building Blocks (ABBs) that perform fundamental mathematical operations—such as addition, \purple{subtraction}, multiplication, and squaring—which are essential for creating splines.
By using these ABBs, we construct a hardware-efficient KAN implementation that reduces area by 125$\times$ and power consumption by 10.59\% compared to a digital spline with 8-bit precision.
Despite introducing an approximation error of up to 7.58\%, our approach provides a promising pathway for more efficient and scalable function approximation in FE. 

\textbf{In summary, the main contributions of this work are:}

\begin{enumerate} 
    \item Creation of ABBs in FE \purple{to} perform \blue{basic} mathematical operations.
    \item \purple{A systematic and generic approach for creating splines and KANs, enabling analog approximation of any given function.}
    \item Comparison of the digital and analog cost of each spline in hardware, \purple{including the physical implementation of the analog spline}.
    \item An analysis of the hardware error and its impact on the overall network \blue{performance}.
\end{enumerate}

The structure of this paper is as follows: ~\autoref{sec:background}, discusses the background.~\autoref{sec:methodology} presents the methodology we followed from the ABB to the approximation of the splines. ~\autoref{sec:Results} analyzes the cost of our implementation and compares it with the digital cost of it, we present as well the error of the approximation and limitations of the work.
Finally,~\autoref{sec:conclusion} concludes the work with future research directions.
\section{Background}\label{sec:background}

\subsection{Flexible Electronics}

FE encompass electronic devices designed to maintain functionality while bending, stretching, or conforming to various surfaces~\cite{FlexibleElectronics}. 
They are typically built on flexible substrates, such as polyimide, plastics, or thin metal foils, enabling them to adapt to a wide range of applications, from wearable health monitors to compact, portable sensors used in consumer and medical electronics~\cite{Heng:AM2022:FlexHumanMachInterfaces,Gao:FlexibleWearableSensing,gao:2016:flexsensor}. 

The manufacturing process for FE \blue{involve} modifications of traditional semiconductor fabrication techniques: 
Instead of rigid substrates, FE are constructed on pliable materials, and certain complex steps—like ion implantation and high-temperature annealing—are often eliminated~\cite{Baruah:FabricationFE2023}.
\blue{For example}, PragmatIC Semiconductor has optimized a streamlined version of silicon lithography specifically for flexible devices and by removing these costly steps, the manufacturing process becomes faster, more cost-effective, and environmentally sustainable due to reduced resource demands~\cite{FlexICs}. 
Furthermore, PragmatIC’s FlexIC technology integrates Thin-Film Transistors (TFTs) on polyimide substrates, achieving critical dimensions of 600 nm. 
This process utilizes a high-k dielectric material that supports standard IC input voltages, enabling compatibility with existing systems while prioritizing ultra-thin and low-energy designs~\cite{FlexICs}.

One of the materials used in FE is Indium Gallium Zinc Oxide (IGZO), an amorphous oxide semiconductor with high electron mobility, transparency, and compatibility with low-temperature processes~\cite{Zhu:IGZO2021,Pan:IGZOTFT2024}. 
These characteristics make IGZO suitable for flexible applications, though its limitations, including its restriction to N-type transistors, impose design constraints~\cite{Jeong:igzoperformance}. 
As such, FE is unlikely to replace silicon-based technologies but to complement them, enhancing hybrid systems where flexibility, low power, and form factor are prioritized over the ultra-precision found in silicon-based circuits~\cite{Lozano:aspdac25:BinCoDesign}.

\blue{FE are well-suited for use cases that require real-time, localized processing in constrained environments, such as monitoring biological signals in wearable health devices or detecting environmental parameters in compact IoT sensors~\cite{Gao:FlexibleWearableSensing,Baruah:FabricationFE2023,Arokia:feAdvantages2012}. 
To meet the specific demands of these applications, FE devices often need efficient function approximation capabilities to perform tasks like signal smoothing, feature extraction, and basic mathematical transformations. 
For instance, wearables may need to compute average heart rate or temperature values, while environmental sensors might use threshold-based approximations for pollution levels or humidity~\cite{Haghi:wearableinhealthmonitoring2021}.

To address the need for efficient, low-power function approximation in FE, we propose an analog approach using KANs, which can effectively model complex, continuous functions in a hardware-efficient manner. 
KANs offer a promising solution for \purple{systematically} implementing these functions in FE, as they allow for real-time analog computation while minimizing the area and power consumption of the system.}

In this work, we use PragmatIC’s FlexLogIC process for rapid, FE production. 
This platform allows for a fast turnaround, typically achieving custom designs in under six weeks~\cite{FlexICs}.
The FlexIC platform provides efficient, pragmatic solutions for FE, especially in applications that benefit from a low-cost, high-volume production model. 

\subsection{Kolmogorov-Arnold Networks}
KANs are inspired \purple{by} Kolmogoro\purple{v}-Arnold representation theorem \cite{theorem}. In~\cite{Liu:KAN2024}, a neural network model is proposed where learnable activation functions are used, unlike \purple{Multilayer Perceptron} (MLP) which has \purple{a} fixed activation function. This replaces the linear weight matrix in MLP with a learnable function dependent on a single variable.
KAN approximates a multivariate continuous function as a composition of continuous functions of a single variable added together. 
To implement these univariate functions, KAN utilizes \purple{Bézier} spline with learnable coefficients~\cite{Liu:KAN2024}. The KAN layer uses the following equation: 
\begin{equation}
\phi(x) = w_b b(x) + w_s \, \text{spline}(x)
\label{KAN}
\end{equation}
where $\phi(x)$ is the sum of basis function b(x) and the spline function. The spline in eq.~\ref{KAN} is a linear combination of \purple{Bézier} splines such that: 
\begin{equation}
\text{spline}(x) = \sum_{i} c_i B_i(x)
\label{eq:spline_sum}
\end{equation}
While the basis function and multiplication can be implemented by multiply and accumulation using a crossbar array \cite{crossbar}, the implementation of spline in analog domain is the focus. The \purple{Bézier} splines have a set of discreet  control points by which the smooth continuous curve is defined. The spline can have n-control points which define the order of the spline. In this work we implement a second-order Bézier \purple{s}pline \cite{secondorderspline} given by: 
\begin{equation}
\mathbf{B}(x) = P_0(1 - x)^2 + 2P_1(1 - x)x + P_2x^2 
\label{eq:bezier_second_order}
\end{equation}
where $P_{0}$, $P_{1}$, $P_{2}$ are the control points of the second order \purple{Bézier s}pline and, \purple{$x$} the input. 
\purple{Choosing a second-order \purple{Bézier} spline reflects a balance between hardware cost and approximation accuracy: while higher-order splines could improve accuracy, they would require more hardware resources, increasing area and power consumption. 
In our case, the second-order spline minimizes hardware and simplifies~\autoref{eq:bezier_second_order} to:}
\begin{equation}
\mathbf{B}(x) = P_0 + (P_1 - P_0)2x + (P_0 - 2P_1 + P_2)x^2
\label{eq:quadratic_formula}
\end{equation}
Eq.~\ref{eq:quadratic_formula} becomes the basis for the implementation of second order \purple{Bézier} splines in the analog domain \purple{using} FE.

\subsection{Related work}

Due to the recent publication of~\cite{Liu:KAN2024}, at the time of this work, only two studies have explored the hardware implementation of KAN.
The first study~\cite{Huang:KANAcceletaror:2024} proposes a mixed analog-digital circuit approach, utilizing Look-up Tables (LUT) and Analog Computing-In-Memory (ACIM). However, LUTs face scalability challenges, as their size grows exponentially with the number of classes, making them memory-intensive.
The second study~\cite{VanDuy:KANCodesign:2024} explores a fully digital circuit implementation\orange{, which also relies on LUTs.}
Both studies incorporate a co-design methodology with software optimization. 
\orange{In this work we do not replicate the designs presented in\cite{Huang:KANAcceletaror:2024} or in\cite{VanDuy:KANCodesign:2024}, as FE require simpler and more resource-efficient designs due to power and area constrains. Instead, to establish a baseline for comparison, we opted for a digital implementation of~\autoref{eq:quadratic_formula} in FE, incorporating the constraints of this technology while excluding the co-optimization strategies used in prior studies.}

Our work presents the first research of a fully analog-domain implementation, leaving the exploration of co-optimization with software for future research. 
\orange{Furthermore, we are the first to explore this design using non-silicon-based technology, which reduces power consumption and requires the creation of simpler and more efficient circuit designs.}
\section{Methodology}\label{sec:methodology}

\blue{
This section describes the methodology employed to implement the KAN in the analog domain, starting with the selection of basic analog components.
These components, referred to as Analog Building Blocks (ABBs), are designed to perform essential operations such as inversion, subtraction, addition, multiplication, and squaring. 
We explain the design choices behind each operation and discuss how they contribute to the overall system's functionality. 
After covering the ABBs, we outline the hardware implementation approach for spline approximation, highlighting how the chosen formula reduces the number of required blocks and enhances both area and power efficiency.
Finally, we present the KAN implementation derived from these splines, \purple{establishing a systematic framework for an analog approximation of any given function.}}

\subsection{Analog Building Blocks}
\begin{figure}[t]
\centering
\includegraphics[width=1\linewidth]{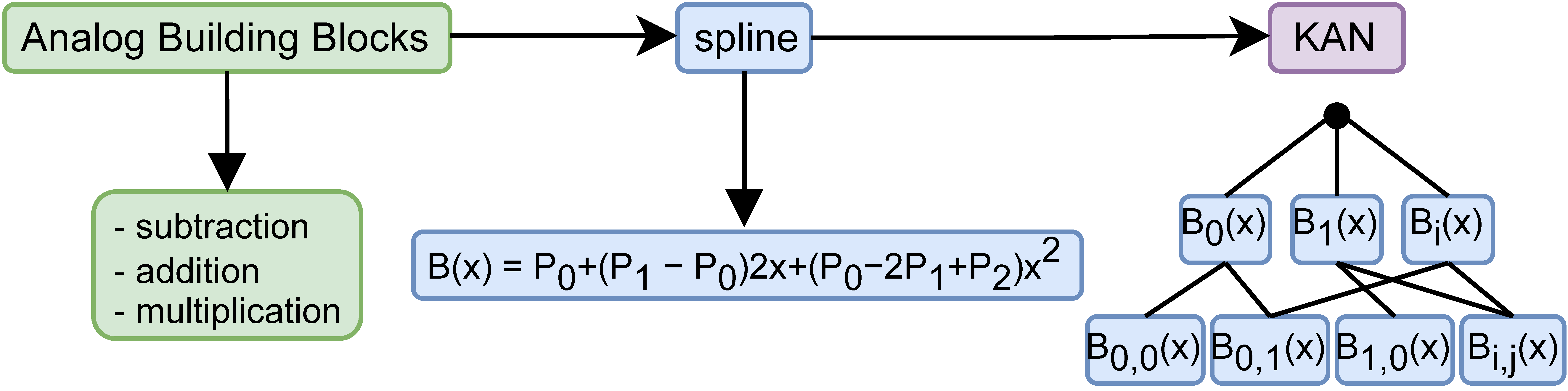}
\vspace{-3ex}
\caption{Proposed methodology to build KAN. The basis for KAN is spline which is interpreted by the second order equation.} 
\label{fig:methodology}
\end{figure}
\begin{figure*}[t]
\centering
\includegraphics[width=1\linewidth]{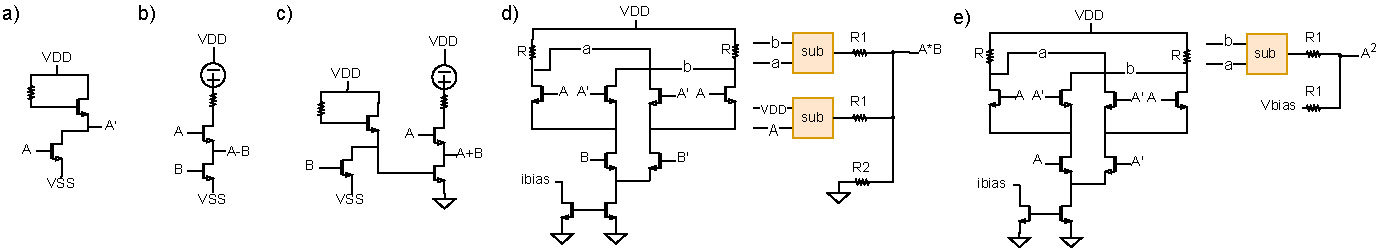}
\caption{Schematic of: a) Inversion b) Subtraction c) Addition d) Multiplication e) Squaring}
\label{fig:schematicABB}
\vspace{-2ex}
\end{figure*}

To implement KAN in the analog domain, we started with basic analog components, which we refer to as "Analog Building Blocks". 
These ABBs perform simple mathematical operations and form the foundation of our design approach, in~\autoref{fig:methodology} we present the methodology followed. 
In this section, we present only the ABBs used to approximate second-degree splines(~\autoref{eq:quadratic_formula}). 
However, we have also developed nonlinear functions, such as softmax, sigmoid, and tanh functions, and more complex operators like integrators, which could be applied in future iterations and other applications.

Due to \blue{large feature sizes in the target flexible IGZO technology which only includes N-type transistors}, we adopted a Resistor-Transistor Logic (RTL) design strategy to minimize the number of necessary components for building blocks. 
Another constraint we faced was the limitation to only N-type transistors, resistors, and capacitors. 
Considering these limitations, \autoref{fig:schematicABB} presents our proposed designs for inversion, subtraction, addition, multiplication, and squaring operations.

\textit{Inversion:} we present a design with only two transistors and one resistor. 
The schematic of the design is shown in~\autoref{fig:schematicABB} a).

\textit{Subtraction:} the design uses two transistors, one resistor, and a bias voltage, $V_{bias}$, which is necessary to expand the operating range. 
The schematic of this design is shown in~\autoref{fig:schematicABB} b).

\textit{Addition:} we employ the logic of $A+B=A-(-B)$. We first use the inversion design on $B$, followed by the subtraction design as $A-(-B)$. 
The schematic of this design is shown in~\autoref{fig:schematicABB} c).

\textit{Multiplication:} we base our design on the Gilbert cell~\cite{Pawase:GilbertCell2018,DIAZSANCHEZ:AnalogMulti:2021}, which provides an optimal solution for analog multiplication in the voltage domain. 
We adapted the design to use only N-type transistors and added a final subtraction $V_{dd}-A$ to improve accuracy across multiple values of $A$. 
The schematic of this design is shown in~\autoref{fig:schematicABB} d).

\textit{Squaring:} we further optimized the multiplication design. 
Knowing the input range in advance allows us to anticipate the output range, simplifying design optimization and reducing area and power consumption. 
We removed the final subtraction block, $V_{dd}-A$, and replaced it with a fixed $V_{bias}$. The schematic of this design is shown in~\autoref{fig:schematicABB} e).

\subsection{Implementation of the spline from multiple ABBs}

\begin{figure}[t]
\centering
\includegraphics[width=1\linewidth]{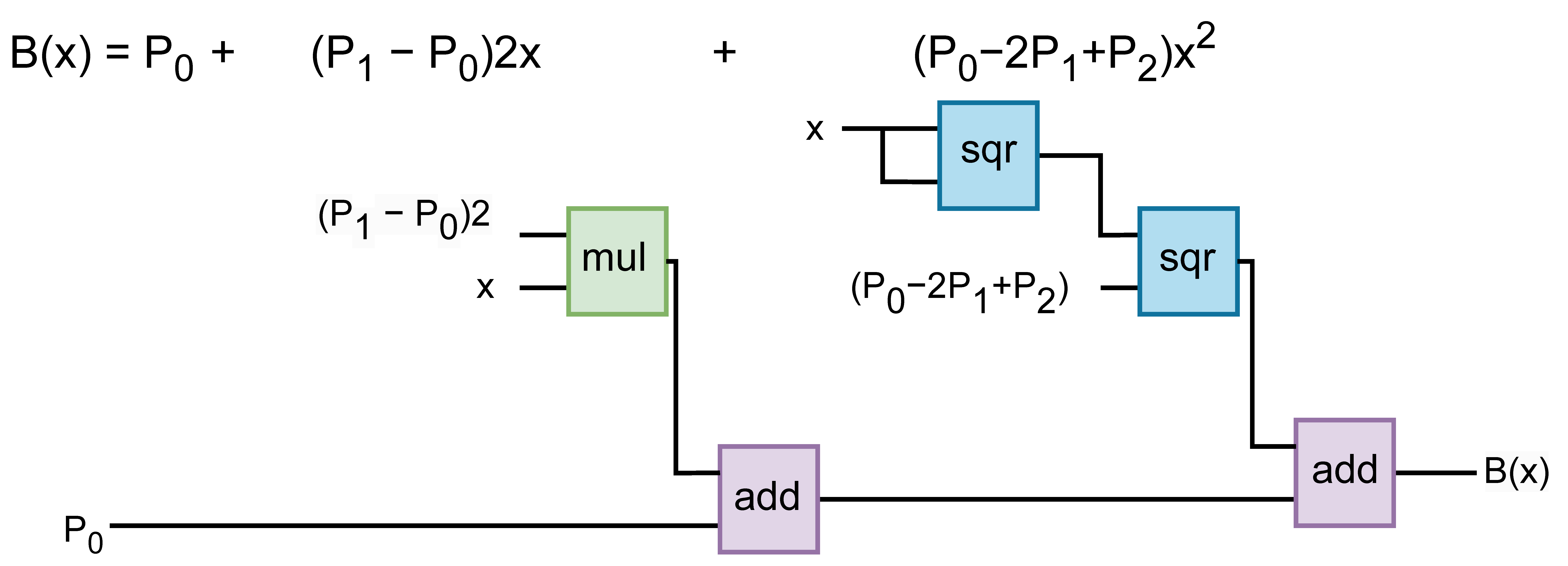}
\caption{Architecture of the spline implementation}
\label{fig:implementation}
\vspace{-2ex}
\end{figure}

For the hardware implementation of each spline, we use the formula presented in~\autoref{eq:quadratic_formula}. 
We chose this equation because it allows us to reduce the number of blocks required, needing only two multiplications, one squaring, and two additions (a total of five blocks). 
In contrast, using~\autoref{eq:bezier_second_order} would require two subtractions, three multiplications, two squaring, and two additions (a total of nine blocks). 
In~\autoref{sec:HardwareCosts}, we present the savings achieved by making this \blue{choice}. 

First, we decompose the expression into three additive sub-blocks: a) $P_{0}$, b) $(P_{1}-P_{0})2x$ and c) $(P_{0}-2P_{1}+P_{2})x^2$.

\begin{itemize}
    \item The first sub-block is simply $P_{0}$, which, as a constant, we interpret as a voltage input. 
    \item The second sub-block consists of $(P_{1}-P_{0})2x$, where $(P_{1}-P_{0})2$ is a constant.
We handle this constant as a voltage input and then multiply it by $x$.
    \item The last sub-block, $(P_{0}-2P_{1}+P_{2})x^2$. Here, $P_{0}-2P_{1}+P_{2}$ is also a constant and will be provided as a voltage input.
To implement this final sub-block, we first square $x$ by multiplying it by itself. We then apply our design shown in~\autoref{fig:schematicABB} e) to perform the multiplication with the constant $(P_{0}-2P_{1}+P_{2})x^2$, which helps to reduce area and power consumption while maintaining precision.
\end{itemize} 
Finally, we sum each of the sub-blocks to construct the spline, resulting in the approximate second-degree function. The full architecture of the implementation is presented in~\autoref{fig:implementation}.

To ensure proper functionality, we carefully designed the input ranges. All ABBs have input ranges from $[-0.5, 0.5]$\orange{, as $[-V_{dd}/2, V_{dd}/2]$}. We need to consider this for the implementation of~\autoref{eq:quadratic_formula}.

For clarity, we examine each of the sub-blocks used to construct the spline, resulting in an approximate second-degree function. Starting with the first sub-block $P_0$, we know there will be no range issues with this term.
For the second sub-block, $(P_1 - P_0) 2x$, we know that $(P_1 - P_0) 2 \in [-0.5, 0.5]$, leading to the first range constraint: 
\begin{equation*}
    P_1 - P_0 \in [-0.25, 0.25]
\end{equation*}
The second range constraint concerns our input $t \in [-0.5, 0.5]$.
For the third sub-block, by reformatting, we obtain:
\begin{equation*}
    P_2 - 2P_1 + P_0 = P_2 - P_1 - (P_1 - P_0)
\end{equation*}

Thus, the range of $P_2 - P_1 - (P_1 - P_0)$ is constrained to $[-0.5, 0.5]$. To determine the range of $P_2 - P_1$, we express it as:
\begin{equation*}
    P_2 - P_1 = (P_2 - 2P_1 + P_0) + (P_1 - P_0)
\end{equation*}

Since the ranges for both $P_2 - 2P_1 + P_0$ and $(P_1 - P_0)$ are known, we can calculate the range of $P_2 - P_1$ by summing the maximum and minimum values of these terms. 

\begin{itemize}
    \item Minimum value of $P_2 - P_1$: 
    \begin{equation*}
        -0.5 + (-0.25) = -0.75
    \end{equation*}
    \item Maximum value of $P_2 - P_1$: 
    \begin{equation*}
        0.5 + 0.25 = 0.75
    \end{equation*}
\end{itemize}

Therefore, we have:
\begin{equation*}
    P_2 - P_1 \in [-0.75, 0.75]
\end{equation*}

Finally, it is important to mention that by considering the trainable values of KAN ($P_{i}$) as inputs in our architecture, the hardware design we present is independent of both the training process and these values. 
Therefore, no redesign is necessary when retraining the model; only the inputs need to be updated. 
For future work, it would be beneficial to perform a co-optimization with fixed $P_{i}$ values, where optimization occurs within each ABB, as was done with the squaring block optimization, based on known input and output ranges. 

\begin{figure}[t]
\centering
\includegraphics[width=1\linewidth]{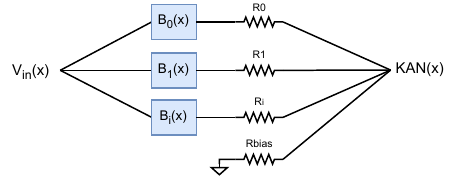}
\caption{Architecture of the KAN implementation}
\label{fig:KAN_from_splines}
\vspace{-2ex}
\end{figure}
\blue{
\subsection{Implementation of the KAN from multiple splines}

Once the splines are created, constructing the KAN requires applying Multiply-Accumulate (MAC) operations.
In analog design, this MAC is achieved by connecting resistors in series with each spline and placing them in parallel to each other.
To implement a weighted MAC, the resistor values can be adjusted according to the desired weighting for each spline.
Alternatively, if uniform weighting is needed, the resistors should have the same value, ensuring each spline contributes equally to the MAC output~\cite{Weller:printedneuron2021}. 
This approach enables flexibility in controlling the contribution of each spline within the KAN structure, aligning with a future co-design. The architecture of the KAN is presented in~\autoref{fig:KAN_from_splines}.
}
\section{Results and Analysis}\label{sec:Results}

\begin{table}
\centering
\caption{Design properties of the analog spline}
\begin{tabularx}{\linewidth}{c|l|X} 
    \noalign{\hrule height 1.2pt} 
    \textbf{Block} & \textbf{Component} & \textbf{Size} \\ 
    \noalign{\hrule height 1.2pt} 
    \multirow{2}{*}{\textbf{INV}} & Transistors & W = 20$\mu$m, L=600nm \\ 
                                  & R & r = 250k$\Omega$, W=2.4$\mu$m, L=3$\mu$m \\ 
    \hline
    \multirow{3}{*}{\textbf{SUB}} & Transistors & W = 20$\mu$m, L=600nm \\ 
                                  & R & r = 30K$\Omega$, W=20$\mu$m, L=3$\mu$m \\ \cline{2-3} 
                                  & \multicolumn{2}{c}{vbias = 0.5V} \\ 
    \hline
    \multirow{5}{*}{\textbf{MUL}} & Transistors & W = 5$\mu$m, L=600nm \\ 
                                  & R & r = 99.97M$\Omega$, W=1.4$\mu$m, L=719.4$\mu$m \\ 
                                  & R1 & r = 428.57k$\Omega$, W=1.4$\mu$m, L=3$\mu$m \\ 
                                  & R2 & r = 40M$\Omega$, W=1.4$\mu$m, L=291.8$\mu$m \\ \cline{2-3} 
                                  & \multicolumn{2}{c}{ibias = 1$\mu$A} \\ 
    \hline
    \multirow{5}{*}{\textbf{SQR}} & Transistors & W = 5$\mu$m, L=1.3$\mu$m \\ 
                                  & R & r = 25.97M$\Omega$, W=1.4$\mu$m, L=201.4$\mu$m \\ 
                                  & R1 & r = 15.45M$\Omega$, W=1.4$\mu$m, L=123.8$\mu$m \\ \cline{2-3} 
                                  & \multicolumn{2}{c}{ibias = 1$\mu$A} \\\cline{2-3}  
                                  & \multicolumn{2}{c}{vbias = -210mV} \\ 
    \cline{1-3} 
    \noalign{\hrule height 1.2pt} 
\end{tabularx}
\label{tab:designsizes}
\end{table}

\subsection{Simulation Setup}

Simulations were made using the Cadence Spectre simulator. For the subtractive flexible technology, we employ the PragmatIC FlexICs PDK second-generation Helvellyn 2.1.0 technology~\cite{FlexICs}. All simulations are performed with $V_{dd}$ of 1.0V, $V_{ss}$ of -1.0V, temperature of 27ºC and our $V_{in}$ is a slope from -0.5V to 0.5V with a rise up time of 10ms.
In~\autoref{tab:designsizes} we present the sizes of each design used in the spline. We do not add the block of the adder, since its design is the result of combining one inversion and one subtraction block.

Since the range of the ABBs inputs was known from the implementation, we were able to optimize each circuit for this specific range. 
For example, the input of the squaring circuit is always between [-0.5,0.5] with an output range of [0,0.25]. 
In contrast, the multiplication circuit's output range can include negative as well as positive values, requiring a more robust circuit. This explains why the squaring circuit has lower resistor values and a higher number of transistors, and why in some cases, is more convenient to use this circuit to perform multiplications instead of the multiplication circuit.

To compare the hardware cost of the analog spline with the digital, we created a standard cell library using PragmatIC PDK \blue{with a $V_{dd}$ of 1V}, synthesized it using Synopsys Design Compiler S-2021.06. 
For simulation and power analysis, we used VCS T-2022.06 and PrimeTime T-2022.03. 
\blue{Based on~\autoref{eq:quadratic_formula} we developed an 8-bit fixed-point digital spline with 6-bit decimal bits and one signed bit,
implemented with sequential logic \purple{using verilog}. The architecture of the digital spline was determined by the synthesis tool.}

\subsection{Hardware costs}\label{sec:HardwareCosts}

\begin{table}
\centering
\caption{Hardware cost of an analog and digital spline}
\scalebox{1}{\begin{tabular}{c|l|l}
    \noalign{\hrule height 1.2pt} 
     \multirow{2}{*}{\textbf{Analog spline}} & area (mm²) & 0.073 \\
    & power ($\mu$W) & 238.5 \\
    \hline
    \multirow{2}{*}{\textbf{Digital spline}} & area (mm²) & 9.111 \\
    & power ($\mu$W) & 266.735 \\
    \noalign{\hrule height 1.2pt} 
\end{tabular}}
\vspace{-2ex}\label{tab:analogvsdigital}
\end{table}

\begin{figure}[t]
\centering
\includegraphics[width=1\linewidth]{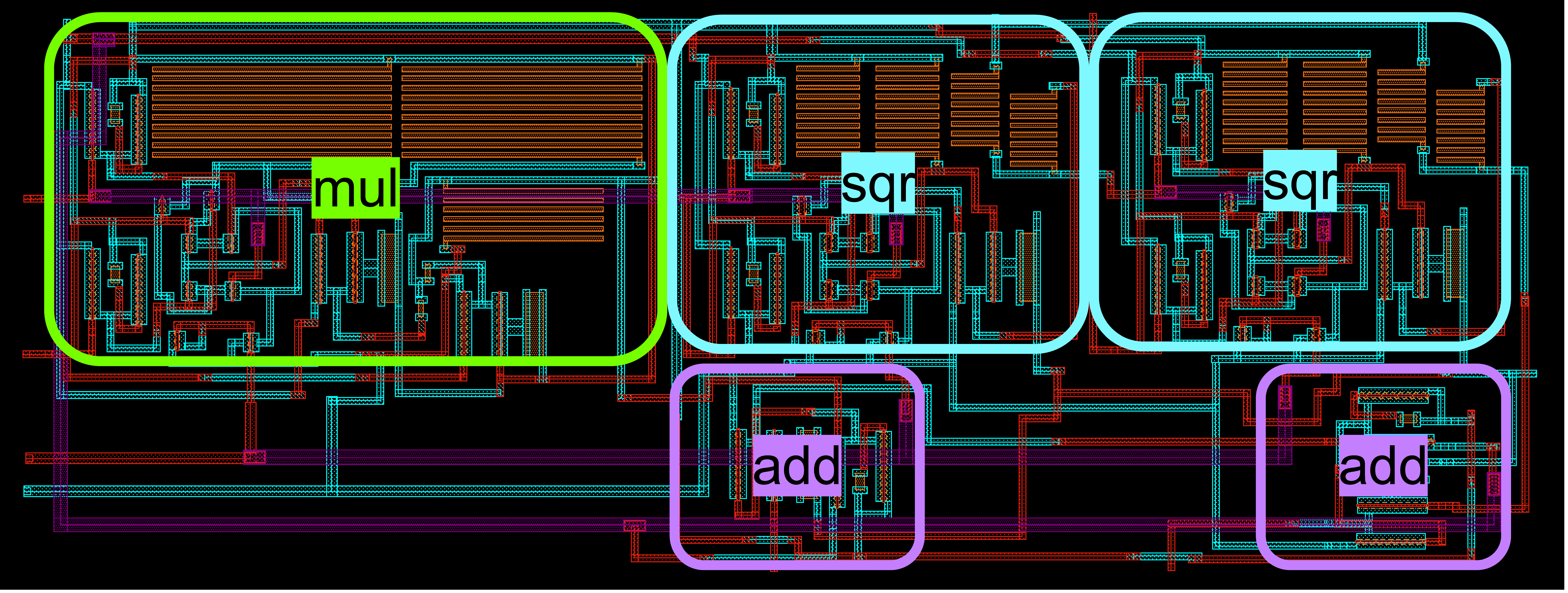}
\caption{Layout of a spline following the architecture presented in~\autoref{fig:implementation}}
\label{fig:splinelayout}
\vspace{-2ex}
\end{figure}

In table~\autoref{tab:analogvsdigital}, we compare the hardware cost of the analog spline presented in this work to \blue{a 8-bit digital spline.}
We can see that the analog implementation's area cost is significantly lower than the digital one, achieving a 125$\times$ reduction. In terms of power, the analog implementation is also more efficient than the digital, with a 10.59\% reduction.
The physical implementation of the analog spline is presented in~\autoref{fig:splinelayout}.
As mentioned in the implementation section, the selection of implementing~\autoref{eq:quadratic_formula} and not~\autoref{eq:bezier_second_order}, reduces area up to 46\% and achieves power savings of 45.7\%.

\subsection{Approximation error}

Since the target of our ABBs is to approximate functions, we expect an error from this approximation. 
We also analyze where the error originates. We use the Normalized Mean Percentage Error (NMPE) metric, \blue{since} it quantifies the average error as a percentage of the function’s range, measuring the magnitude and direction of errors. 
This relative measure is particularly useful for evaluating function approximation within normalized signal ranges.

\textit{Error from each ABB}: We first measured the error in each block, as presented in~\autoref{tab:errorABBs}.
A negative NMPE indicates that the approximation tends to underestimate the function, while a positive value suggests that the approximation overestimates it.

\begin{table}
\centering
\caption{Error of each ABB}
\scalebox{1}{\begin{tabular}{l|c} 
    \noalign{\hrule height 1.2pt} 
    \textbf{Block} & \textbf{Normalized Mean Percentage Error (NMPE)} \\ 
    \noalign{\hrule height 1.2pt} 
    \textbf{INV} & 2.66\% \\ 
    \textbf{SUB} & -4.7\% \\ 
    \textbf{ADD} & -4.23\% \\ 
    \textbf{MUL} & -6.99\% \\ 
    \textbf{SQR} & -7.93\% \\ 
    \noalign{\hrule height 1.2pt} 
\end{tabular}}
\vspace{-2ex}\label{tab:errorABBs}
\end{table}

\textit{Error of the full spline}: When we integrated all our ABBs together, we observed that the error is not commutative, but in many cases, it compensates. 
We ran three different scenarios with predefined values for $P_{0}$, $P_{1}$ and $P_{2}$, and the maximum error was -7.58\%\orange{, observed in Scenario C.
The results for each scenario, including the values of $P_{0}$, $P_{1}$ and $P_{2}$, along with the corresponding NMPE, are summarized in~\autoref{tab:errorFullSpline}.}.

\begin{table}
\centering
\caption{Error Analysis of the Full Spline Across Different Scenarios}
\scalebox{1}{\newcolumntype{Y}{>{\centering\arraybackslash}X}

\begin{tabularx}{\linewidth}{c|lcY}
    \noalign{\hrule height 1.2pt} 
    \textbf{Scenario} & \textbf{Parameter ($P_i$)} & \textbf{Value} & \textbf{NMPE (\%)} \\
    \noalign{\hrule height 1.2pt} 
    \multirow{3}{*}{\textbf{A}} & $P_0$ & 0.1 &\\
                                & $P_1$ & 0.3 &-4.84\\
                                & $P_2$ & 0.8 &\\
    \hline
    \multirow{3}{*}{\textbf{B}} & $P_0$ & 0.2 &\\
                                & $P_1$ & 0.4 &-5.02\\
                                & $P_2$ & 0.6 &\\
    \hline
    \multirow{3}{*}{\textbf{C}} & $P_0$ & 0.15 &\\
                                & $P_1$ & 0.4 &-7.58\\
                                & $P_2$ & 0.85 &\\
    \noalign{\hrule height 1.2pt} 
\end{tabularx}}
\label{tab:errorFullSpline}
\end{table}

\begin{figure}[t]
\centering
\includegraphics[width=0.95\linewidth]{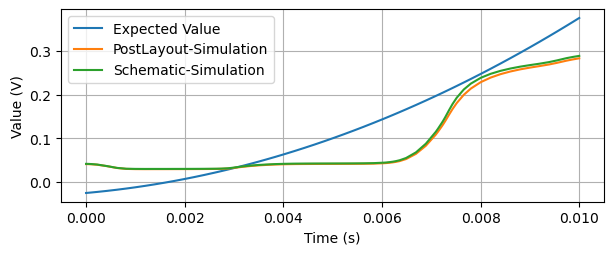}
\caption{Comparison of error between schematic and post-layout simulations. The plot shows the impact of parasitic elements and layout errors on the model's accuracy. \red{The 'Expected Value' represents the mathematical result obtained from Eq. 4.}} \label{fig:schevslayout}
\vspace{-2ex}
\end{figure}

\textit{Error of the \blue{physical implementation}}: We also need to consider the error introduced by the layout and its parasitic elements. 
In the example with the most error, the NMPE from the schematic simulation was -6.34\% but with the post-layout simulations, this error increased to -7.58\% (\orange{value presented in~\autoref{tab:errorFullSpline}}). 
A comparison between the schematic and post-layout simulations can be seen in~\autoref{fig:schevslayout}.

\begin{figure}[t]
\centering
\includegraphics[width=1\linewidth]{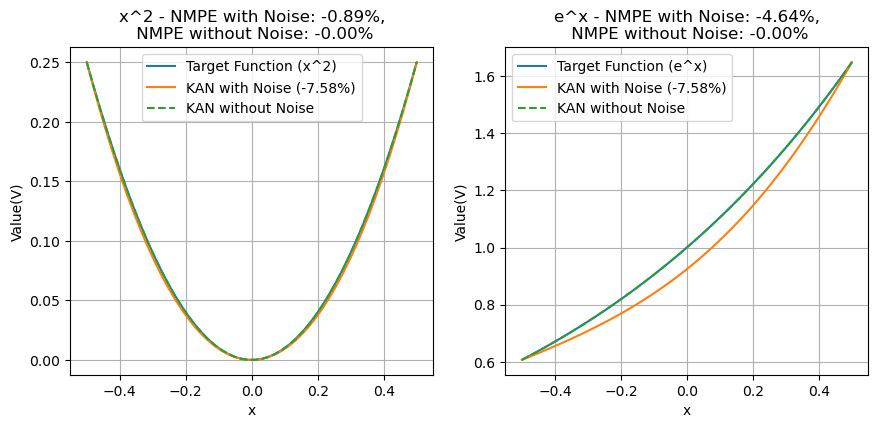}
\caption{Comparison of KAN approximations for two target functions, $x^2$ and $e^x$, with and without the hardware error. The corresponding Normalized Mean Percentage Errors (NMPE) are included for both cases.}\label{fig:KANerror}
\vspace{-2ex}
\end{figure}

\textit{Impact of Hardware-Induced Error on Function Approximation Using KAN}: To investigate the sensitivity of KAN-based function approximation to hardware errors, we developed a Python script using numpy, matplotlib, and scipy CubicSpline. 
This script simulates a simplified KAN model with one input and multiple basis splines. 
The network was trained to approximate two chosen target functions, $x^2$ and $e^x$, using a fixed range of input values $[-0.5,0.5]$.

The KAN implementation features three basis splines, each represented by a cubic spline. 
A controlled noise factor of -7.58\% was applied to each spline during training to emulate hardware error, introducing a stochastic perturbation that mirrors potential inaccuracies in hardware approximations. 
For comparison, we ran a parallel experiment without noise to observe the baseline performance of the KAN.
The output of these simulations is shown in~\autoref{fig:KANerror}. As we can see, the effect of the hardware error is minimal: \orange{less than 1\% for $x^2$ and less than 5\% for $e^x$}.
It is also important to note that expanding the input range to higher values, such as [-1,1] or [-5,5], significantly reduces the NMPE to below 1\% in the error-induced KAN in both cases.

\blue{\subsection{Discussion of the results}}

This work has several limitations that suggest promising directions for future research. 
First, our focus is limited to approximating second-degree functions, which restricts the range of applications. 
Extending this approach to higher-order functions would increase its applicability. 
\red{For higher precision requirements or more complex function approximations, it would be necessary to either use more splines—thus increasing the KAN network—or to improve and add more ABBs.}

Since this approach relies on approximation, it naturally introduces some degree of error. 
\red{While these results demonstrate the resilience of KAN-based approximation to hardware errors, minimizing hardware error requires a robust design that maintains simplicity and is specifically tailored to our FE implementation.
This presents a trade-off between error reduction and preserving the hardware benefits of analog design. Balancing these factors is critical in ensuring an optimal approximation with minimal complexity.}

Another limitation is that we did not incorporate co-design strategies, which might have enabled more optimized and efficient configurations\red{, increasing accuracy and versatility, as seen in~\cite{Huang:KANAcceletaror:2024,VanDuy:KANCodesign:2024}}.
Lastly, our design operates in the voltage domain, which limits the operating range. 
Future studies might explore shifting to the current domain to extend this range and adapt to broader application needs.

These results highlight the benefits of analog computing for target applications such as ABBs and the implementation of splines to create KAN. Nevertheless, it is important to mention that while the objective of this work is to create an analog function approximation, which provides benefits in area and power savings, it comes at the cost of higher error compared to the digital equivalent. Therefore, digital computing may be a better solution when high precision is mandatory. 

\vspace{2ex}
\section{Conclusion}\label{sec:conclusion}

Flexible Electronics (FE), while flexible and lightweight, impose challenges such as limited conductivity and larger footprints. 
Analog approximations offer a promising solution by reducing both power and area.
This paper explores the potential of analog Kolmogorov-Arnold Networks (KANs) for function approximation in FE, \purple{presenting a systematic and generic design process that enables approximations for any given function.}
By developing Analog Building Blocks (ABBs) for basic operations such as addition, multiplication, and squaring, we created splines that serve as the foundation of the KAN.
Our results show significant hardware improvements, including a 125$\times$ reduction in area and a 10.59\% decrease in power consumption compared to a digital 8-bit spline implementation. 
Although the analog design introduces an approximation error of up to 7.58\%, it does not substantially degrade the overall performance of the system.
This analog approach offers considerable advantages, particularly in applications where power efficiency and area optimization are prioritized over precision.

{\small
\section*{Acknowledgment}
This work is partially supported by the European Research Council (ERC), and co-funded by the H.F.R.I call “Basic research Financing (Horizontal support of all Sciences)” under the National Recovery and Resilience Plan “Greece 2.0” (H.F.R.I. Project Number: 17048).
}

\bibliographystyle{IEEEtran}
\bibliography{reference}

\end{document}